\documentclass{aa}
\usepackage{graphicx}
\begin{document}
   \title{Spectroscopic Observations of Twenty-one Faint Cataclysmic Variables  Candidates\thanks{Based on observations obtained at ESO La~Silla (ESO Proposal 69.D-0142(A))}}

   \author{E. Mason 
		\inst{1} \and 
	   S. B. Howell 
		\inst{2}
	   }

\offprints{E. Mason}

   \institute{ESO, Alonso de Cordova 3017, Casilla 19001, Vitacura, 
		Santiago, Chile \\
              \email{emason@eso.org}
         \and
             Institute for Geophysics and Planetary Physics, University of California, Riverside, CA\\}

\abstract{
We provide the first minimum light spectroscopic observations 
for 21 previously known or suspected faint cataclysmic variable candidates. 
The sources were selected from the Downes et al. (2001) living edition catalog
and the identified candidates have minimum light magnitudes of V$\sim$18-22.
We confirm 15 of the candidates to be cataclysmic variables. 
 
}

   \titlerunning{Faint CVs}
   \authorrunning{E. Mason \& S. B. Howell}
   \maketitle
%

\section{Introduction}

Cataclysmic variables are a subclass of interacting binary star which contains a white dwarf primary  and, typically, a low mass late type secondary star.  
The white dwarf accretes mass from the secondary star. Accretion is the cause of modulated brightness behaviors with the dwarf novae class showing semi-periodic
outbursts. 
Among dwarf novae the subclass of WZ~Sge-like stars or TOADs (Tremendous Outburst Amplitude Dwarf novae, Howell et al. 1995) shows very infrequent large amplitude outbursts and is believed to represent the oldest, most evolved CV population. There is observational evidence for very low mass ($\leq$0.05 $M_\odot$) brown dwarf-like secondary stars in few of these objects (e.g. Harrison et al. 2002, Howell \& Ciardi 2001).  Recent theoretical work (i.e. Howell et al. 2001) suggests that the vast majority of CVs at present day should be of very short orbital period, intrinsically faint systems, having properties similar to the TOADs. In general, very faint CVs are discovered during outburst, after which they fade into obscurity. Spectroscopy and other information {\bf are} rarely available for them at minimum light and many have only ill-defined positions with some faint star listed as the likely candidate.
Thus, the search for and spectroscopic confirmation of additional faint CVs is  important as it provides a statistically larger sample of such highly evolved systems. Study of such systems will test and improve our understanding of the evolution of binary stars containing compact objects.

A list of previously known or suspected cataclysmic variables was collected from the CV catalog of Downes et al. (2001) based on the lack of spectral information at minimum light.
Our selected sample had candidate stars of V magnitude from 18 to fainter then 22, some not even having a positive candidate identification. 
Fifteen, out of the twenty-one candidates we observed, turned out to be bona fide CVs.

We present below our observational information and the list of the target stars
as well as a brief discussion for each object. 

\section{Observations}

Our observations were performed at the ESO NTT (New Technology Telescope) plus EMMI (ESO Multi-Mode Instrument) on the night 10 August 2002. 
We used EMMI in RILD (Red Imaging Low Dispersion) mode with a 360 l/mm grism (grism \#3) which provided 2.3 \AA \ spectral resolution and wavelength coverage from $\sim$3800 to $\sim$9200 \AA.
The seeing was between 0.4 to 0.6 arcsec for all objects observed and
we used a 1.0 arcsec slit. 
Spectro-photometric standards were observed at the start and 
near the end of the night and bias, flat, and wavelength calibration arcs were obtained prior to and after our night of observing. 
Our calibrated spectra have a wavelength scale uncertainty or RMS of $\sim$ 2 \AA \ and typical flux uncertainties are of order of 15\%.
A complete description of our used EMMI setup is given in Howell et al. (2002).

\begin{table}
\begin{center}
\scriptsize
\caption{Log of observations and object V magnitude. Objects missing the V band magnitude correspond to misidentified CVs.}
\begin{tabular}{cccc}
 & & & \\
UT start  &  Int Time(sec) & Source & V \\
 & & & \\
 23:24:33.0  & 1800 & AB Nor  & 20.3 \\
 00:05:14.0  &  900 & FL TrA   & -  \\
 00:29:09.0  & 1800 & V592 Her & 21.3 \\
 01:08:55.0  & 1200 & V544 Her & 19.5  \\
 01:38:32.0  & 900  & MM Sco   & 17.9  \\
 02:18:58.0  & 1000 & YY Tel   & 18.6  \\
 02:44:29.0  & 900  & V2359 Sgr&  - \\
 03:09:43.0  & 1000 & V551 Sgr & 19.5  \\
 03:31:16.0  & 1000 & V662 CrA & 16.0  \\
 03:59:52.0  & 1000 & V344 Pav & 19.4  \\
 04:24:22.0  & 750  & DH Aql   & 17.0  \\
 04:43:21.0  & 1000 & V1233 Aql& 18.7  \\
 05:06:57.0  & 750  & V1141 Aql& 19.4  \\
 05:26:09.0  & 750  & V1047 Aql& 18.3  \\
 05:50:15.0  & 750  & AW Sge   &  -  \\
 06:15:50.0  & 650  & V2276 Sgr&  -  \\
 06:33:30.0  & 750  & IS Del   & 18.3  \\
 06:52:08.0  & 3600 & AO Oct   & 21.0  \\
 08:25:07.0  & 400  & BE Oct   & 20.4  \\
 08:54:36.0  & 900  & VW Tuc   & 19.4  \\
 02:49:40.0  & 1000 & X Cir    & 19.4  \\
 & & & \\ 
\end{tabular}   
\end{center}
\end{table}

\section{Faint CV Candidate Sources}

Table 1 presents the log of the observations and a list of the known or suspected CVs we observed. The coordinates can be obtained from the catalog of Downes et al. (2001), thus, are not repeated herein. Table~1 also lists the object V magnitude (column~4) estimated by convolving our observed spectrum with a standard V-Johnson filter. 
 We present below a brief discussion for each object and discuss some of the saline features. 
Table~2 lists line and continuum flux measurements plus EW for the Balmer, HeI and FeII~42 emission lines, common to almost all of our targets.  

\subsection{AB Nor}

\begin{figure*}
\rotatebox{-90}{\includegraphics[width=14cm]{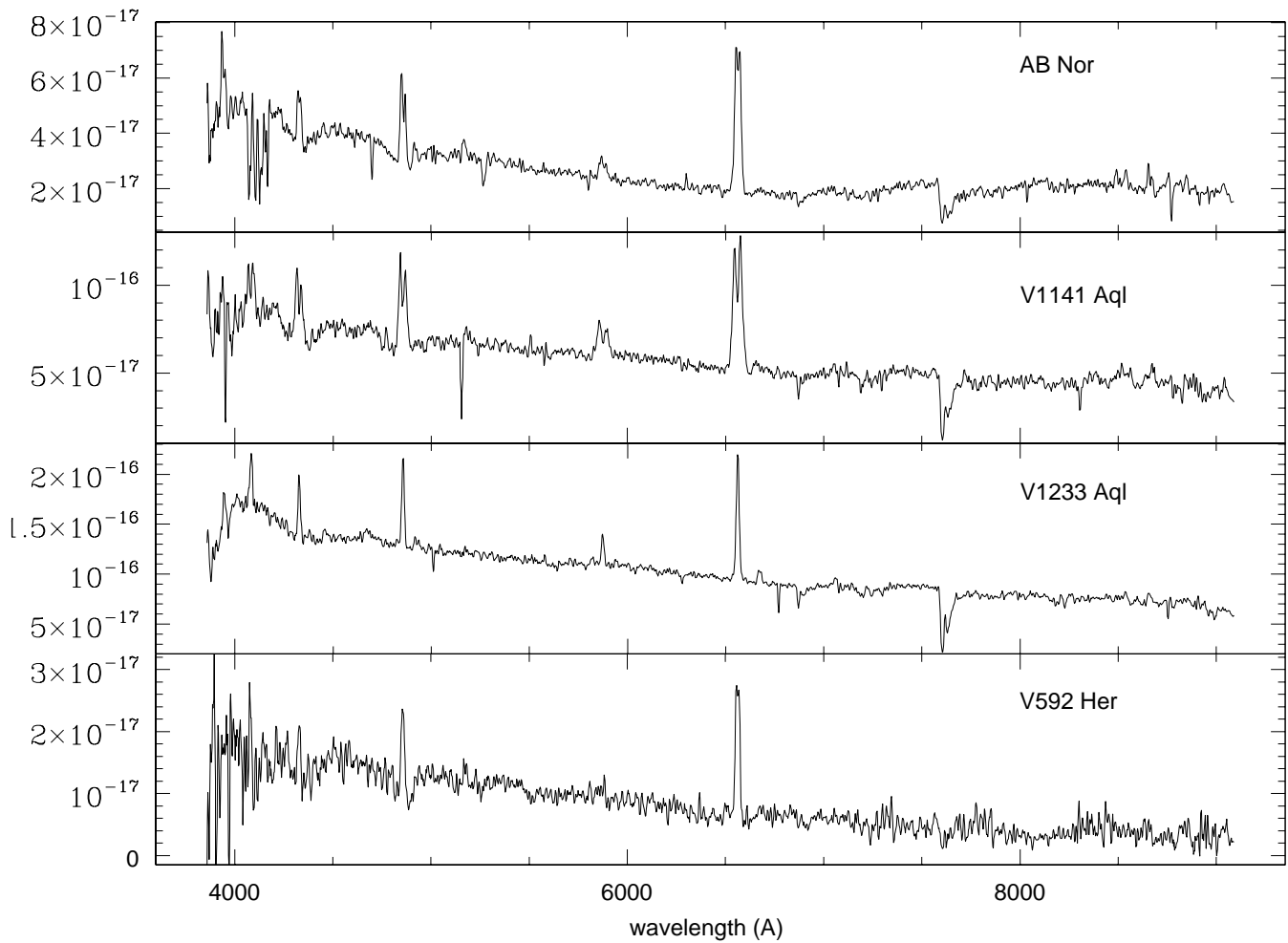}}
\caption{Spectra of AB~Nor, V1141~Aql, V1233~Aql, and V592~Her. They are all systems showing blue continuum and H line emission, HeI and heavier elements being weak or ``lost in the noise''. Flux units are in erg/sec/cm$^2$/\AA. }
\end{figure*}

AB Nor is classified as a SU UMa star in the Downes et al. living edition CV catalog (Downes et al. 2001). 
The object has been apparently observed in superoutburst and showed superhumps which allowed the derivation of the orbital period (P$_{orb}\sim$0.076 days, Downes et al. 2001). Before 1982 observations of AB~Nor were reported only by Bateson (1982a and 1982b) who failed in catching any outburst as the object magnitude 
was always below the telescopes limit magnitude (typically in the range 13.5-14.2)
The VSNET database reports detection of 2 or 3 outbursts since April 2000\footnote{Observed outbursts occurred on April 2000 and September 2002. Due to the broken nature of the observations is not clear whether the April 2000 outburst is just a long single one or two short, which are only $\sim$5 days apart.}. The average magnitude of the outbursts in the VSNET database is $\sim$14.0. During the September 2002 outburst, B. Monard and P. Nelson possibly detected growing superhumps and determined a likely superhump period of 0.0826 days.

Our AB Nor spectrum (Fig.~1), shows that the object is a CV with double peaked Balmer lines (FWHM of the H$\alpha$ emission line is $\sim$1700 km/sec), and evidence for white dwarf Balmer absorptions (H$\alpha$ excluded). HeI emission lines are also present ($\lambda$ 4471 and 5876\AA), as well as FeII (multiplet 42: $\lambda\lambda$ 4924, 5018, and 5169 \AA, while multiplet 38 is weaker). At longer wavelengths are the H(P) lines 9 and 10.   
The continuum is blue but rises again beyond 7000 \AA, possibly implying evidence for the secondary star contribution. 

\subsection{V1141 Aql}

V1141~Aql is classified  as UG type CV with no outburst detected by Downes et al. (2001).  However, the VSNET database has reported about 5 outbursts  since 1999 (on October 1999, August 2000, July 2001, July 2002, and September 2002). The average maximum magnitude is 15.2 with the exception of the July 2001 outburst which brightened to magnitude 14.7 nine days after the outburst began. 

This object is characterized by large double peaked emission lines (FWHM of H$\alpha$ is $\sim$2500 km/sec), which are signature for a high orbital inclination (see Fig.~1). The HeI line 5876 is evident, while all other HeI emissions are lost in the continuum noise. The continuum rises blue-ward and shows weak evidence for white dwarf absorptions at H$\beta$ and H$\gamma$.

\subsection{V1233 Aql}

Similar to V1141~Aql, V1233~Aql is classified as an UG type CV by Downes et al. (2001). No outburst have been reported and detected in Downes et al. catalog, nor in the VSNET database. However, the object is said to vary between 15.5 and 20 mag (photographic) by Downes et al. (2001). 

V1233 Aql (Fig.~1) shows a blue continuum and very narrow  Balmer and HeI~5876 \AA \  emission lines (FWHM of H$\alpha \sim$800 km/sec). Weak are the HeI lines 4471, 6678, 7065 and possibly the HeII 4686. Weak are also the Paschen series lines, while uncertain is the identification of the FeII~42 multiplet (blending with HeI emission lines).  From our spectroscopic observation V1233~Aql was about $\geq$1 mag brighter than the quiescent magnitude reported in Dowens et al. (2001).

\subsection{V592 Her}

V592 Her was found in outburst in 1968 (Richer 1968), 1986 (Richer 1991), and 1998 (Waagen et al. 1998). Waagen et al. observed the object at a magnitude V=12 (which implies an outburst amplitude A $>8$ mag). 

Classified as a CV of uncertain nature in the GCVS, V592~Her was selected as a candidate halo CV by Howell \& Szkody (1990) and observed in time resolved photometry by Howell et al. (1991). They report an average quiescent magnitude of V=20.5 ($\pm$0.1) and no periodic modulations but just strong flickering ($\sim$0.3-0.4 mag). V592~Her was classified as a TOAD of unknown orbital period by Howell et al. (1995). 
The third outburst was followed photometrically by Duerbeck \& Mennickent (1998) who report the detection of superhumps and provide a likely orbital period of $\sim$0.06 day. Duerbeck \& Mennickent observations definitively confirm V592~Her to be an SU~UMa star, or better a TOAD, considering its long recurrence time and outburst amplitude. Light curve of the 1998 outburst is in the VNSET database. 

V592 Her spectrum (Fig.~1) presents narrow (H$\alpha$ FWHM$\simeq$1200 Km/sec) Balmer emission lines. Emission lines from HeI and heavier elements appear to be present but are possibly lost in the noise. The continuum emission is weak and raising blue-ward. Weak evidence of the underlying white dwarf appears in the Balmer absorption blue-ward of the H$\alpha$ line. 

\subsection{AO~Oct}

\begin{figure*}
\rotatebox{-90}{\includegraphics[width=14cm]{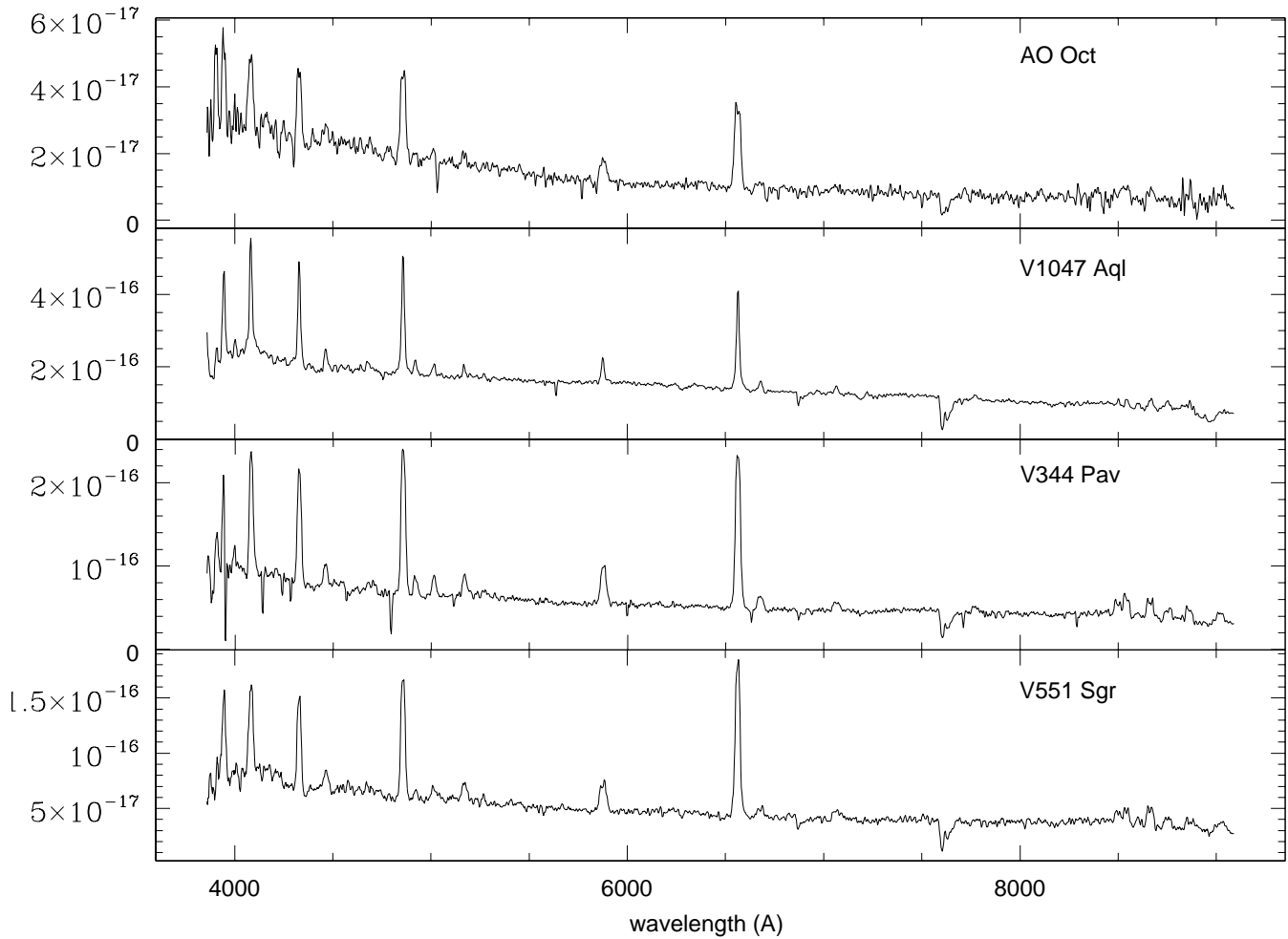}}
\caption{ Spectra of AO~Oct, V1047~Aql, V344~Pav, and V551~Sgr. They are characterized by slightly blue continuum and relatively strong HeI, FeII and H-Paschen emission lines. Flux units are in erg/cm$^2$/sec\AA. }
\end{figure*}

AO~Oct is classified as a WZ~Sge type dwarf nova based on its outburst amplitude (A=7.5 mag according to the mag range of 13.5-21.0 in Downes et al. 2001). However, the VSNET database reports no outburst brighter than 14.2\footnote{The detected outbursts occurred on June 1999, September 2000, plus possibly three more on June 1997, March 1998 and September 1998} (though not well covered). 

Quiescent photometric observations were reported by Howell et al. (1991) who failed in detecting any significant modulation and, later, by Woudt \& Warner (2002) who determined an average quiescent magnitude of V=20 and an orbital period of 0.0654 days. 
 
Our AO~Oct spectrum if Fig.~2 shows relatively broad Balmer lines (H$\alpha$ FWHM $\sim$ 1700 km/sec but only marginal evidence of double peaked profiles. 
HeI lines are almost lost in the continuum noise with the exception of the HeI~11 at 5876 \AA \ which is relatively strong. However, we detected also the HeI~14, HeI~46, HeI~10, and probably the FeII~42 multiplet (blending with HeI). H(P) 9 lines are visible at the red end of the spectrum. AO~Oct's continuum raises toward the blue. 

\subsection{V1047 Aql}

V1047~Aql lacks any outburst report both in the Downes et al. catalog (2001) and the VSNET database, similarly to V1233~Aql. Downes et al. (2001) classify V1047~Aql as an UG type star showing brightness variation in the magnitude range (photographic) 15 to less than 17.5.

V1047~Aql spectrum (Fig.~2), shows a blue continuum, very strong but very  narrow Balmer emission lines (H$\alpha$ FWHM $\sim$730 km/sec) and inverse Balmer decrement (H$\beta >$ H$\alpha$ and possibly H$\delta > $ H$\gamma$). Relatively strong are the FeII~42 lines, the HeI 4471 and 5876, while weaker the HeI lines 6678 and 7065 and the H Paschen series. 
 
\subsection{V344 Pav}

V344~Pav was discovered in outburst by Maza \& Hamuy (1990) on the 21st of July 1990, at a magnitude of M$_{pg}$=14.5. 
Five days later the object had spectroscopic follow-up (Phillips 1990), which revealed V344~Pav to be a dwarf nova in outburst. An archive search of the Cerro Roble Observatory plates (covering the period May 1979 - September 1984, Barros 1990) showed that V344~Pav was in outburst also on September 1983 and March 1984 when the object was at magnitude M$_{pg}$=14.5 and 16, respectively. Quiescent magnitude was estimated to be $\ge$20. 

The VSNET database reports about 4 observed outbursts since 1999, occurred respectively on September 1999, July and September 2000, and May 2001. 
Both the 1999 and 2001 outbursts were caught during rise and reached maximum magnitude of 14.3. 

The spectrum of V344~Pav (Fig.~2) shows a slightly blue continuum and strong emission lines from H and HeI. Clearly visible are both the Balmer and the Paschen hydrogen series, the FeII~42 multiplet, and the HeI lines 7065, 6678, 5876, 4713, 4471, and the higher excitation line HeII 4686. V344~Pav shows evidence for an inverse Balmer decrement being H$\delta >$ H$\gamma$, in integrated flux (see Table~2).
The FWHM of the H$\alpha$ emission line is $\sim$1300 km/sec. 

\subsection{V551 Sgr}

V551~Sgr was observed by Bateson (1982a), who reports the detection of three outbursts among 142 observations taken between 1978 and 1981. The average magnitude of the outbursts is 13.5 but the observations are too sparse to allow distinction between short and long outbursts. 

The VNSET database reports three more outbursts occurred in October 1998, April 2000, and May 2001. Their average magnitude is 13.9-14.0 (May 2001 outburst was probably caught during decline), in good agreement with Bateson observations. The reported outburst activity and the observed outburst amplitude (A$\ge$5.6) is possibly consistent with a WZ~Sge like object (or TOAD).

V551~Sgr spectrum in Fig.~2 shows strong Balmer lines, relatively strong HeI emission and slightly blue continuum,  similarly to V1047~Aql and V344~Pav, presented above.  Clearly evident are also the H Paschen series and the FeII, multiplet 42. The continuum raises blue-ward. H$\alpha$ FWHM ($\sim$ 1200 km/sec) is consistent with a low orbital inclination system (similarly to V592~Her), and there is no evidence of double peaked profiles other than in the Paschen series H lines. 

\subsection{BE~Oct}

\begin{figure*}
\rotatebox{-90}{\includegraphics[width=14cm]{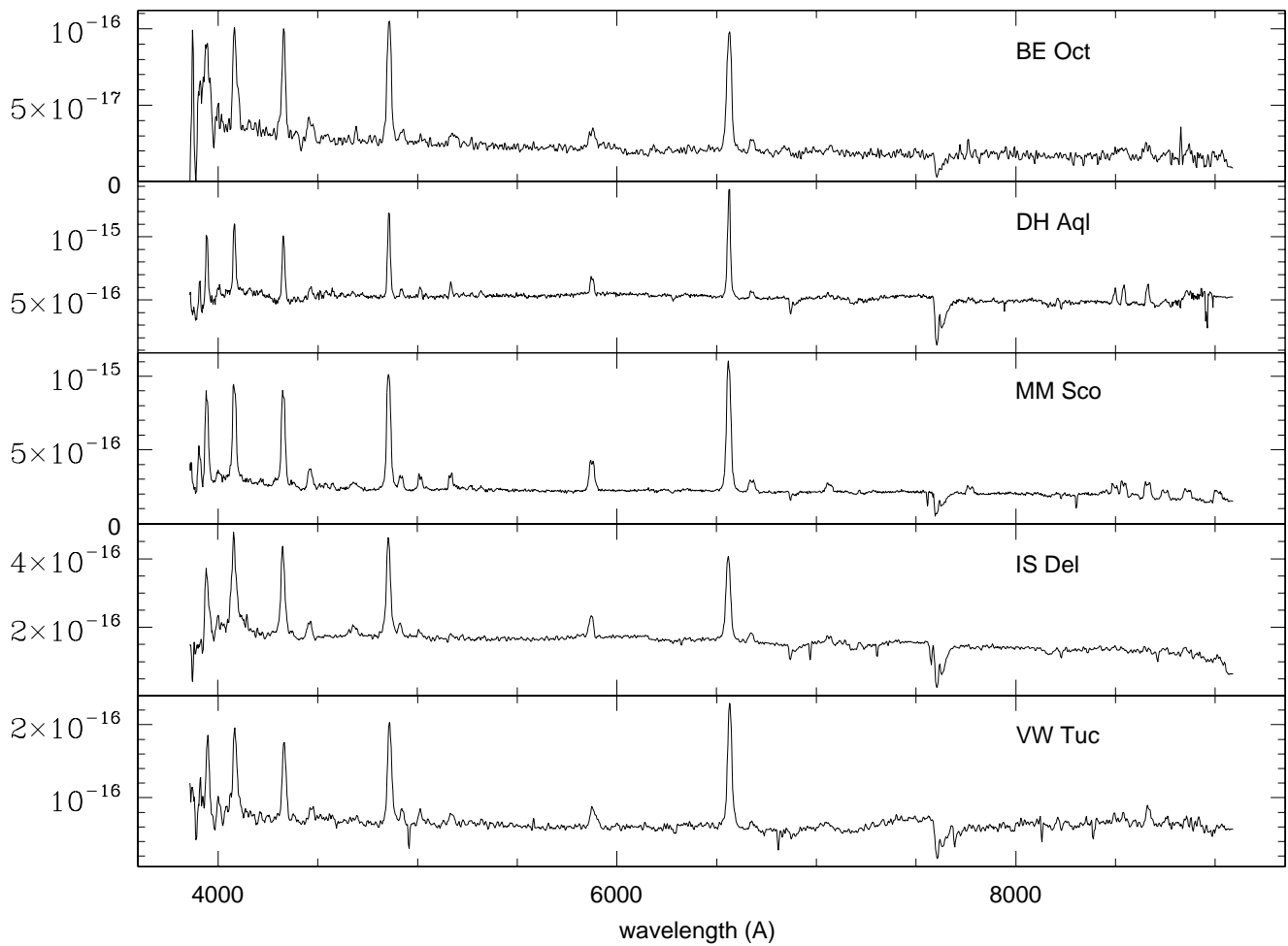}}
\caption{Spectra of BE~Oct, DH~Aql, MM~Sco, IS~Del, VW~Tuc. These 5 systems are all characterized by flat continuum and relatively strong HeI and H-Paschen lines, as well as FeII emission lines. Flux units are erg/cm$^2$/sec/\AA. }
\end{figure*}

BE~Oct was observed at fainter magnitude ($\sim$1 mag, see Table~1) than reported by Downes et al. (2001). The only observed outburst in the literature is the August 1998 one, during which the object was observed in fast speed photometry by Kemp and Patterson (VSNET alert 513). They report a maximum magnitude of V=15.94 and detection of 0.23 mag superhump modulation 1-2 days after maximum. The determined superhump period is of 0.07712$\pm$0.00013 days.

The spectrum (Fig.~3) slightly rises blue-ward in the continuum, while emission lines are from H, HeI (4471, 5876, 6678, and 7065), FeII~42, and HeII (4686). Unidentified is the emission at $\sim$6848\AA. Paschen lines are also visible (H(P) 10 and 9) at the red end of the spectrum. The H$\alpha$ FWHM is of $\sim$1300 km/sec.

\subsection{DH Aql}

DH~Aql was discovered by Tsessevich (1969) who analyzed Sterngberg plates finding three outbursts. These outbursts were short lived and with rapid declines. Tsessevich  derived an outburst recurrence time of 268 days.
Another outburst was reported in 1972 by Zhukov and Solovjev (1972). 
The object was later monitored somehow regularly by Bateson (1982a, 1982b, 1984) who observed DH~Aql 94 times over 2 years (from April 1980 to May 1982). Bateson caught the star at V=13.4 only once, in August 1980. The monitoring was then continued by Jones, Brown and Stubbings until September 1997, and a total of 951 observations were eventually analyzed by Bateson (1998) who reports 10 detected outbursts, 7 of which were super-outbursts. Bateson (1998) also provides an average super-maxima magnitude of 12.6 and an average super-cycle length of 361 days, though he notices that super-cycle can be as short as 259 as well as long as 450 days. Normal outburst have average magnitude of 14.0 but their recurrence time could not be determined due to the broken nature of the observations (Bateson 1998). Bateson (1998) concludes DH~Aql to be a SU~UMa star. 

Confirmation of the SU~UMa nature of DH~Aql arrived by Nogami \& Kato  (1995) who observed DH~Aql in time resolved spectroscopy just after a VSNET alert, and estimate the object at magnitude V=12.4. They detected superhumps and determined a superhump period of 0.0805 days. 

The VSNET database reports 11 outbursts between 1995 and 2002\footnote{According to the VSNET database the outbursts occurred on July 1995, May, July, and October 1997, July and October 1998, April and July 1999, August 2000, August 2001, and July 2002}, though mostly poorly monitored with just a few points near maximum. 
Magnitudes at maximum are consistent with those provided by Bateson. 

DH Aql (Fig.~3) show very narrow H emission lines (H$\alpha$ FWHM $\leq 800$km/sec), as well as He~I (lines 14, 11, and 46), and FeII (multiplet 42). Strong are the Paschen series H lines (10 and 9), series 10 possibly blending with CaII emission (8498 and 8542). The continuum emission is flat with evidence of white dwarf absorption at H$\gamma$ and blue-ward. 

\subsection{MM~Sco}

MM~Sco was monitored over several years by Bateson (1982a, 1982b, 1984), starting since 1959. Bateson's data correspond to those MM~Sco outbursts whose brightness was above the telescopes limit magnitude (typically around 14-15 magnitude). He reports detection of 12 outbursts having average magnitude of 13.6. Four of the 12 detected outbursts were observed to persist for more than 6 days with one of them lasting about 30 days, three were classified as narrow (i.e. short outburst); while, the outburst duration of the remaining five is uncertain. 
On April 1984 Bateson (1984) detected one more outburst and confirmed that ``wide'' outbursts (i.e. of longer duration) are about 0.5 mag brighter than the ``narrow'' ones. 
Bateson et al. (1997), combine previous and new MM~Sco outburst detections and conclude that MM~Sco alternates wide and narrow outbursts having average magnitude of 13.6. Bateson et al. compute an average inter-outburst period of  65.7 days (though very variable, i.e. ranging between 10 and 166 days), but do not exclude  much shorter recurrence time, should smaller outbursts of magnitude $\sim$15 (i.e. below the telescope limit) occur. 

The VSNET database reports detection of 5 outbursts since July 1998 (July 1998, July 1999, July 2000, April 2001, and September 2002). During the September 2002 outburst MM~Sco was observed in time-series photometry by B. Monard who reports detection of 0.1 mag superhumps having a likely period of 0.062 days (VSNET alert 7468).   

MM~Sco was brighter than quoted by Downes et al. (2001), at the time of our observation (see Table~1). Fig.~3 shows that MM~Sco is characterized by flat continuum and very strong hydrogen lines (both Balmer and Paschen series, in the optical and near IR respectively). Strong are also the He lines 4471 (HeI 14), 5876 (HeI 11), 6678 (HeI 46), 7065 (HeI 10) and the Fe~II 42. Also in emission are HeII~4686 (HeII~1, blending), FeII~38 and 49, OI~4 (7772), and possibly CaII~2 (8498 and 8542, blending with H(P)10 lines). All emission lines but the Balmer series appear double peaked. The H$\alpha$ Balmer emission line has FWHM=1100 km/sec.

\subsection{IS Del}

The object is listed as a UG type CV in the Downes et al. catalog (2001) typically showing $\sim$2.5 mag brightness variations (mag range 15 to $\leq$17.5). The VSNET database reports about 4-5 outbursts poorly monitored and probably not observed at maximum as the data points range between magnitude 15.3 and 16. 

IS~Del spectrum (Fig.~3) is characterized by strong Balmer emission lines over a relatively flat continuum. Balmer lines show an inverse Balmer decrement similar to V1047 Aql (H$\beta > $H$\alpha$). The FWHM of the H$\alpha$ is $\sim$1200 km/sec. IS Del also shows HeI emission lines (14, 11, 46 and 10), and the HeII 4686 (blending). The hydrogen Paschen series shows only weak emissions. 

\subsection{VW~Tuc}

Classified as an uncertain UG CV in the GCVS, was proposed as halo CV candidate by  Howell \& Szkody 1990. Hoard \& Watcher (1998) observed VW~Tuc both in broad band photometry and low resolution spectroscopy and confirmed it to be a CV. According to Hoard \& Watcher's photometric observations, WV~Tuc is a 15.4 mag object, much brighter than derived by us in Table~1, but in good agreement with the magnitude range provided by Downes et al. (2001). Hoard and Watcher's spectrum covers the wavelength range 5750-7250\AA \ and shows H$\alpha$ and HeI emissions, plus a very blue continuum. 

Our spectrum in Fig.~3 shows a much flatter continuum and strong ``modulations'' above 7000\AA \ which are likely indicative of the secondary star's contribution. 
Strong narrow Balmer lines (H$\alpha$ FWHM $\sim$1100 km/sec), and an inverse Balmer decrement (H$\delta > $H$\gamma$), characterize the spectrum of VW~Tuc in Fig.~3. The visible wavelength range is also characterized by HeI and FeII emission lines, while the HeII 4686 is quite weak. Line profiles are all single peaked and with relatively small FWHM (H$\alpha$ FWHM $\sim$1100 km/sec) both indicative of a low orbital inclination CV. Hoard \& Watcher derived an orbital inclination of only 20$^o$ from the analysis of the H$\alpha$ line profile. 

\subsection{V544 Her}

\begin{figure*}
\rotatebox{-90}{\includegraphics[width=14cm]{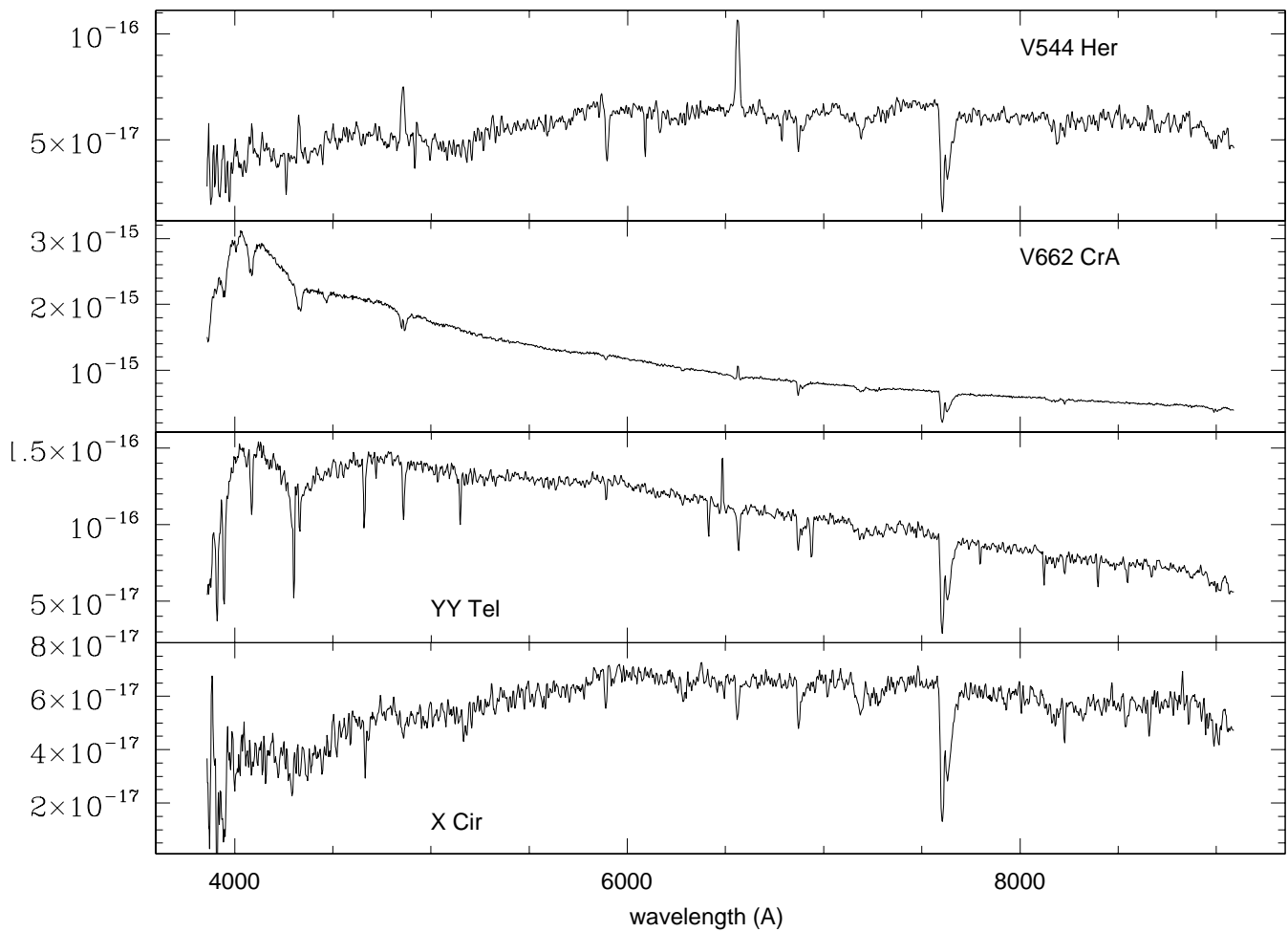}}
\caption{Spectra of V544~Her, V662~CrA, YY~Tel, and X~Cir. V544~Her shows and ``anomalous'' continuum, while V662~CrA likely corresponds to an outburst spectrum. YY~Tel remains a systems of uncertain classification. X~Cir is likely misidentified. See test at corresponding paragraph for more details. Flux units are ergs/cm$^2$,sec/\AA. }
\end{figure*}

V544~Her was first proposed to be a CV by Hoffmeister (1967) who observed variations of $\ge$5.5 magnitudes. Howell et al. (1990) observed V544~Her during their faint CVs photometric survey and found an orbital period of 0.069($\pm$0.0035) days. 
According to Vanmunster \& Howell (1996), V544~Her was in outburst on September 1994, reaching a max magnitude of 14.5 (photographic). While, Spogli et al. (1998) caught the object during decline from an outburst on July 1995  determining its B-V, V-R, and V-I colors. 
In the VSNET data base there is evidence for a third outburst in October 2000. The database shows the object at magnitude V=16, but we cannot say whether it was a smaller outburst or just observation during the declining phase of an outburst. 
 
The spectrum of V544 Her (Fig.~4) shows very narrow Balmer emission lines (H$\alpha$ FWHM $\sim$900 km/sec) and an extremely modulated continuum with absorption bands probable signatures of the secondary star. 
The combination of narrow accretion disk emission lines together with the modulated continuum may imply that the system is in a very low $\dot{M}$ state (hibernation), and/or is nearly face-on, and/or a long orbital period system.

\subsection{V662 CrA}

Classified as V$^\star$ in the GCVS, V662 CrA appears as a UG type CV in Downes et al. (2001) catalog. No photometry, nor spectra either in quiescence or in outburst appear in the literature, but according to Downes et al. (2001), V622~CrA shows brightness variation in the range 15.7 to $<$19.6. 

Our V662~CrA spectrum (Fig.~4) is peculiar as it shows only H$\alpha$ in emission, while bluer Balmer lines present absorption superimposed to a weak emission core (the emission does not reach the continuum level). 
The continuum emission is very blue. Fig.~4 probably corresponds to  maximum spectrum as the object was found brighter by $\sim$ 3 mag than expected. 
Moreover, our computed magnitude V=16 (see Table~1) is consistent with the maximum observed magnitude reported by Downes et al. (2001). Thus, this is likely the first outburst detected for V662~CrA. 

\subsection{YY Tel}

YY~Tel is classified as V$^\star$ in the GCVS, while Vogt \& Bateson (1982) classified it as possible WZ~Sge type star lacking of any photometric and spectroscopic confirmation either in quiescence or in outburst. Finally, YY~Tel is a UGWZ type CV in Downes et al. catalog (2001). Downes et al. (2001) do not report  any outburst detection but magnitude variations in the range of 14.4-19.3. No outburst observations of YY~Tel appear anywhere in the literature. 

The first YY~Tel spectrum was obtained by Zwitter \& Munari (1996) who also provided continuum measurements and UBVR$_C$I$_C$ photometry. Zwitter \& Munari did not observe any emission line but just a hot blue continuum. 

Recent observations of YY~Tel have been carried out by Woudt \& Warner (2001) who started a photometric survey of faint CVs. They observed the object in two runs but did not detect flickering and/or orbital modulations. However, they observe 0.15 mag variations on a time scale of few days during their second run. They report YY~Tel colors similar to those of an F-type sub-dwarf. 

Our spectrum in Fig.~4 is similar to the one presented by Zwitter \& Munari as it also lacks of emission lines\footnote{Though we possibly see an emission at $\sim$ 6482 \AA.}. However, our observed continuum appears to be fainter and with a stronger blue-ward rise with respect to the Zwitter \& Munari's spectrum. The classification of the object remain uncertain, but it does not appear to be a CV. 
 
\subsection{X~Cir}

Downes et al. (2001) classify X~Cir as a Nb CV. It was observed in outburst in 1927 both photometrically and spectroscopically (Becker 1929). The maximum observed magnitude was 6.5 (photographic, Bode \& Evans 1989) and no further observations have been performed since then. Quiescent observation always failed due to the extreme faintness of any candidate star, which made its identification uncertain. However, X~Cir was recently re-identified by Woudt and Warner (private communication) who detected strong flickering but no orbital modulation during 1/2 hr observations (Woudt \& Warner 2002). We observed X~Cir according to Woudt and Warner's identification and the spectrum is shown in Fig.~4. Clearly, it does not present typical signatures of a cataclysmic variable and likely the object has probably been misidentified once more.

\begin{table*}
\begin{center}
\scriptsize
\caption{Spectroscopic measurements. Line and continuum fluxes are in erg/cm$^2$/sec/\AA. Line measurements correspond to the integrated flux. Continuum measurements correspond to the average of the cursor position marked on each wing of the emission line.}
\begin{tabular}{lcccccccc}
 & & & & & \\
Object & H$\delta$ & H$\gamma$ & H$\beta$ & H$\alpha$ & HeI~14 & HeI~11 & HeI~46 & FeII~42  \\
 &line & line &line &line &line &line &line &line \\
 &cont & cont &cont &cont &cont &cont &cont &cont \\
 & EW &EW &EW &EW &EW &EW &EW &EW \\
 & & & & & & & & \\
AB Nor   &  & 4.935E-16 & 9.166E-16  & 1.995E-15 & 9.245E-17  & 2.662E-16b & &1.689E-16 -- 1.631E-16 \\
 & &3.602E-17 &2.836E-17 &1.835E-17 &3.854E-17 &2.370E-17 & & 2.836E-17 -- 3.073E-17\\
 & &-13.82 &-32.2 &-108.8 & -2.399 & -11.22 & &-5.932 -- -5.306 \\
V592 Her & 1.837E-16  & 1.673E-16 & 3.465E-16 & 6.252E-16 & & & &\\
 & 1.299E-17 &1.200E-17 &9.085E-18 &5.449E-18 & & & &\\
 & -14.18 &-13.98 &-38.35 &-114.8 & & & &\\
V544 Her &  & 3.160E-16 & 6.002E-16 & 1.029E-15 & & & &\\
 & &4.116E-17 &4.831E-17 &6.353E-17 & & & &\\
 & &-7.678 &-12.43 &-16.2 & & & &\\
MM Sco   & 1.659E-14 & 1.619E-14 & 2.085E-14 & 2.498E-14 & 3.929E-15 &6.678E-15 &2.799E-15 &2.276E-15 2.461E-15 2.772E-15 \\
 & 2.941E-16 &2.610E-16 & 2.338E-16 & 2.184E-16 &2.338E-16 &2.216E-16 &2.142E-16 &2.309E-16 2.272E-16 2.302E-16\\
 &-56.48 &-62.04 & -89.12 & -114.5 &-16.8 &-30.12 &-13.08 &-9.841 -10.82 -12.04\\
V551 Sgr & 2.181E-15 & 2.155E-15 & 3.017E-15 & 4.378E-15 &6.750E-16 &1.080E-15 &4.013E-16 &1.634E-16 3.894E-16 5.150E-16 \\
 & 8.025E-17 &6.505E-17 &5.829E-17 &4.173E-17 &6.399E-17 &4.709E-17 & 4.133E-17&5.839E-17 5.797E-17 5.712E-17\\
 & -27.18 &-33.25 &-51.75 &-104.9 &-10.55 &-22.94 &-9.699 &-2.8 -6.718 -9.013 \\
V344 Pav & 3.601E-15 & 3.395E-15 & 4.787E-15  & 5.902E-15 &6.960E-16 &1.655E-15 &5.751E-16 &5.123E-16 6.022E-16 7.604E-16\\
 &8.897E-17 & 7.916E-17 & 6.699E-17 &4.859E-17 &7.585E-17 &5.437E-17 &4.721E-17 &6.494E-17 6.384E-17 6.399E-17\\
 &-40.37 & -42.91 &-71.51 &-121.6 & -9.181 &-30.47 &-12.19 &-7.87 -9.426 -11.87 \\
DH Aql   & 9.459E-15  & 9.749E-15 &1.158E-14  & 1.663E-14 &2.085E-15 &2.894E-15 &1.414E-15 &9.995E-16 1.241E-15 1.740E-15\\
 & 5.563E-16 &4.814E-16 &5.248E-16 & 5.169E-16&5.078E-16 &5.472E-16 &5.112E-16 &5.248E-16 5.223E-16 5.248E-16\\
 &-17.0 &-20.26 &-22.06 & -32.19&-4.107 &-5.289 &-2.767 &-1.904 -2.376 -3.314\\
V1233 Aql& 8.681E-16 & 9.786E-16 & 1.506E-15 &2.460E-15  &3.133E-16  &4.750E-16 &2.971E-16 &\\
 & 1.691E-16 &1.358E-16 &1.284E-16 &9.372E-17 &1.334E-16 &1.097E-16 &9.100E-17 &\\
 & -5.137 &-7.218 &-11.75 &-26.25 &-2.349 &-4.33 &-3.265 &\\
V1141 Aql& 1.418E-15 &1.262E-15 & 2.004E-15 & 3.626E-15 & &1.003E-15 &1.203E-16? &\\
 &7.816E-17 &6.985E-17 &6.288E-17 &5.162E-17 & &5.968E-17 &5.154E-17 &\\
 &-18.15 &-18.06 &-31.87 &-70.25 & &-16.81 &-2.334 &\\
V1047 Aql& 6.918E-15 & 5.780E-15 & 6.427E-15 & 5.643E-15 &1.269E-15 &1.314E-15 &6.215E-16 &7.016E-16 6.300E-16 5.214E-16\\
 & 2.328E-16 &2.013E-16 &1.823E-16 &1.375E-16 &1.895E-16 &1.555E-16 &1.329E-16 &1.795E-16 1.754E-16 1.712E-16 \\
 & -29.74 &-28.93 &-35.26 &-41. &-6.691 &-8.452 &-4.684 &-3.904 -3.594 -3.045\\
IS Del   & 4.945E-15 & 6.383E-15 & 7.905E-15 & 7.237E-15 & & & &\\
 & 1.612E-16 &1.814E-16 &1.759E-16 &1.638E-16 & & & &\\
 & -30.74 &-35.26 &-44.95 &-44.22 & & & &\\
AO Oct   & 6.348E-16 & 6.570E-16 &9.376E-16  & 9.653E-16 &1.829E-16 &3.107E-16b &1.007E-16 &9.426E-17 - 9.828E-17 \\
 &2.730E-17 &2.224E-17 & 1.710E-17 & 9.394E-18 &2.254E-17 &1.091E-17 &8.903E-18 &1.684E-17 - 1.631E-17\\
 &-23.25 &-29.52 & -54.81 &-102.8 &-8.114 &-28.46 &-11.31 &-5.599 - -6.03\\
BE Oct   & 1.619E-15 & 1.584E-15 & 2.195E-15 & 2.387E-15 &4.336E-16 &4.675E-16 &2.402E-16 &2.361E-16 - 3.049E-16b \\
 & 3.587E-17 &3.072E-17 &2.557E-17 & 2.067E-17 &2.742E-17 &2.185E-17 &1.966E-17 &2.550E-17 - 2.414E-17\\
 & -45.07 &-51.61 &-85.86 &-115.3 &-15.78 &-21.4 &-12.19 &-9.266 - -12.63\\
VW Tuc   & 3.583E-15 &2.543E-15  &3.744E-15  & 4.875E-15 &5.037E-16 &9.695E-16 &2.119E-16 &4.006E-16 4.193E-16 4.551E-16\\
 & 7.393E-17 &7.128E-17 &6.494E-17 &5.922E-17 &6.877E-17 &5.966E-17 &5.827E-17 &6.479E-17 6.435E-17 6.317E-17\\
 & -48.49 &-35.73 &-57.67 &-82.33 &-7.311 &-16.25 & -3.638 &-6.182 -6.516 -7.206\\
 & & & & & & & &\\ 
\end{tabular}   
\end{center}
\end{table*}

\subsection{Mis-identified CVs}

V2276~Sgr, V2359~Sgr, FL~TrA, and  AW~Sge were clearly misidentified in the Downes et al. (2001) catalog. The observed spectra correspond to G-type stars in the case of V2359~Sgr, FL~TrA, and  AW~Sge, and a hot (possibly DC) white dwarf in the case of V2276~Sgr. There is obviously a need to re-identify these four objects. 

\section{Conclusions}
We observed in low resolution spectroscopy 21 faint CVs or suspected CVs which lack quiescent spectrum in Donwes et al. catalog (2001). Among the 21 objects:
\begin{itemize}
\item four (V2276~Sgr, V2359~Sgr, FL~TrA, and  AW~Sge), resulted misidentified in Downes et al. catalog finding charts,
\item one, X~Cir, is still missing proper identification due to its extremely faint quiescent magnitude 
\item one, YY~Tel remains with an uncertain classification
\item the remaining 15 objects have their CV nature confirmed spectroscopically.
\end{itemize}

Most of the identified CVs appear to be low orbital inclination system having 
narrow Balmer emissions (FWHM $\leq$1200 km/sec). Clear double peaked profile in the emission lines have been detected only in V1141~Aql and AB~Nor, which likely are high and mid orbital inclination system, respectively. AO~Oct may have double peaked lines as well but will be only a mid-low orbital inclination system (see Fig.~2). 

Four over fifteen systems ($\sim$27\%) have a very blue continuum, strong Balmer emission lines, and much weaker or no HeI emissions. Five over fifteen objects (i.e. 30\%) show flat coninuum and strong Balmer, HeI, and FeII emission lines. Four systems are somehow of intermediate type between the two described above, as they show slightly blue continuum and relatively strong emission lines from HeI and FeII. There is evidence for an inverse Balmer decrement in V1047~Aql, VW~Tuc, IS~Del, and V344~Pav. An inverse Balmer decrement implies larger temperatures, which could either depend on the location of the emission lines forming gas or on the mass trasfer rate, or both. 
We count 3 (out of 15) TOAD candidates (AB~Nor, V551~Sgr, and IS~Del) based on their outburst amplitude ($\geq 6$ mag). V344~Pav likely is a TOAD too, should its quiescent magnitude be confirmed fainter than 20 (rather than 19.4 as at the time of our observing run). V592~Her is a TOAD based on its large outburst amplitude, the rareness of the observed outburst and the detection of superhumps during a superoutburst. Superhumps detection in AB~Nor is presented as possible and not as certain. 
V544~Her, DH~Aql and BE~Oct, which have detected superhumps and superoutburst, are SU~UMa star due to their smaller outburst amplitude. 
MM~Sco is a candidate SU~UMa star based on the alternance of large and small outburst, though super humps have not been observed yet. 
We can conclude that 9 (out of 15 confirmed CVs, i.e. 60\%) are likely low mass transfer rate short orbital period systems, below the period gap.

\begin{acknowledgements}
VSNET data base for having data public available.      
    
\end{acknowledgements}

\end{document}